\documentclass[11pt,twoside]{article} % Specifies the document style.
\usepackage{times,fancyhdr}
\usepackage{cite}
\usepackage{graphicx}
\date{October 27, 2006}
\title{Computers and Liquid State Statistical Mechanics} % Declares the document's title.
\author{Carl McBride\\
Instituto de Qu\'{\i}mica F\'{\i}sica Rocasolano (CSIC),\\ Serrano 119,\\ 28006 Madrid,\\ Spain
} % Declares the author's name.
\begin{document} % End of preamble and beginning of text.
\pagestyle{fancy}
\fancyhead{} % clear all header fields
\fancyhead[EC]{Carl McBride}
\fancyhead[EL,OR]{\thepage}
\fancyhead[OC]{New Research on Computer Physics}
\fancyfoot{} % clear all footer fields
\renewcommand\headrulewidth{0.5pt}
\addtolength{\headheight}{2pt} % make space for the rule
\maketitle % Produces the title.
The advent of electronic computers has revolutionised the application of statistical mechanics to the liquid state.
Computers have permitted, for example, the calculation of the phase diagram of water and ice
and the folding of proteins.
The behaviour of alkanes adsorbed in zeolites, 
the formation of liquid crystal phases and the process of nucleation.
Computer simulations provide, on one hand, new insights into the physical processes in action, 
and on the other,  quantitative results of greater and greater precision.
Insights into physical processes facilitate the reductionist agenda of physics, whilst
large scale simulations bring out emergent features that are inherent (although far from 
obvious) in complex systems consisting of many bodies.
It is safe to say that computer simulations are now an indispensable tool for both the
theorist and the experimentalist, and in the future their usefulness will only increase.

This chapter presents a selective review of some of the incredible advances in condensed matter physics
that could only have been achieved with the use of computers.
%%%%%%%%%%%%%%%%%%%%%%%%%%%%%%%%%%%%%%%%%%%%%%%%%%%%%%%%%%%%%%%%%%%%%%%%%%%%%%%%%%%%%%%%%%%%%%%%%%%%%%%%%%%%%%%%%%
\section{Introduction}
Most mechanical problems that have an element of realism are analytically intractable;
a famous example being the motion of three or more mutually attractive bodies.
However,  almost all of these problems are suseptible to computation, leading to approximate solutions of arbitrary accuracy.
Likewise, very few problems in statistical mechanics are exactly solvable \cite{e_book_exactly_solved}.
It is because of this that computational methods are so invaluable.
%%%%%%%%%%%%%%%%%%%%%%%%%%%%%%%%%%%%%%%%%%%%%%%%%%%%%%%%%%%%%%%%%%%%%%%%%%%%%%%%%%%%%%%%%%%%%%%%%%%%%%%%%%%%%%%%%%
\section{Statistical mechanics of liquids before computers}
The study of liquids is a field that has benefitted greatly from advances in computational power. 
Initial theoretical approaches to the study of liquids did not meet with the same level
of progress as did the study of weakly interacting gasses and crystalline solids. A great part of the problem lies in the fact that
there is no simple, analytically tractable, idealised model to start from. Liquids incorporate some configurational structure, 
as demonstrated by their radial distribution functions, but they also have a dynamical aspect typified by gasses, 
i.e. diffusion.
Thus the treatment of liquids as either high temperature solids, or as dense gasses, met with little success. 

Other factors have helped to stymie an analytical development of the theory of liquids, 
such as the failure \cite{PR_1952_085_000777} of the, reasonable sounding, Kirkwood superposition approximation \cite{JCP_1935_03_00300}.
This approximation treated a three-body correlation as being the result of three two-body correlations:
\begin{equation}
{\rm g}_N^{(3)}({\bf r}_1,{\bf r}_2,{\bf r}_3)={\rm g}_N^{(2)}({\bf r}_1,{\bf r}_2){\rm g}_N^{(2)}({\bf r}_2,{\bf r}_3){\rm g}_N^{(2)}({\bf r}_3,{\bf r}_1),
\end{equation}
where ${\rm g}_N^{(n)}$ is the $n$-particle distribution function.
This approximation was central to the Born, Bogoliubov, Green, Kirkwood, and Yvon (BBGKY) hierarchy of integral equations. 
The BBGKY relations are exact, but closed solutions could now only be obtained for pair-wise additive systems. 
However, the importance of three-body forces on the gas-liquid equilibrium has been highlighted by Anta {\it et al.} \cite{PRE_1997_55_002707}.
Also, in a description of binary liquids, the use of accurate two-body potentials must be supplemented by
three-body terms in order to obtain realistic results \cite{JCP_2006_125_074503}. 
%%%%%%%%%%%%%%%%%%%%%%%%%%%%%%%%%%%%%%%%%%%%%%%%%%%%%%%%%%%%%%%%%%%%%%%%%%%%%%%%%%%%%%%%%%%%%%%%%%%%%%%%%%%%%%%%%%
\subsection{Liquids as a dense gas}
For an ideal gas one has the famous `ideal gas law',
\begin{equation}
P=\frac{Nk_BT}{V} = \frac{ \rho}{\beta},
\end{equation}
where $P$ is the pressure, $N$ is the number of molecules, $k_B$ is the Boltzmann constant, $T$ is the temperature, and $V$ is the volume.
$\rho$ is the number density, $N/V$, and $\beta$ is defined to be $1/k_BT$.
In an attempt to describe a liquid as non-ideal gas one can expand the ideal gas expression as 
\begin{equation}
P=\frac{Nk_BT}{V} \left(1 + \frac{N B_2(T)}{V} +  \frac{N^2 B_3(T)}{V^2} + ... \right),
\end{equation}
where $B_n(T)$ is known as the $n^{\rm th}$ virial coefficient
(such an  expression can be  justified via a diagrammatic Mayer expansion of the grand partition function in irreducible clusters \cite{PTP_1961_025_0537}).
With a knowledge of the virial coefficients one is able to then calculate the fugacity, internal energy, enthalpy, molar heat, entropy, and the Joule-Thompson coefficients \cite{bookDymond_new}.\\
The second virial coefficient is given by
\begin{equation}
B_2(T) = - \frac{1}{2} \int f(r) d{\bf r},
\end{equation}
and the third is given by
\begin{equation}
B_3(T) = - \frac{1}{3} \int \int f(r) f(r') f(|{\bf r}-{\bf r}'|)d{\bf r}d{\bf r}',
\end{equation}
where $f(r)$ is the Mayer $f$-function, defined as $f(r)= \exp [-\beta \Phi(r)] -1$, and $\Phi(r)$ is the interaction potential.
As can be seen, the expressions for the virial coefficients become increasingly involved.
For example, for one of the simplest systems, the three dimensional hard sphere fluid, only $B_2$, $B_3$ and $B_4$ 
have been derived analytically \cite{KNAW_1899_1_0273,KNAW_1899_1_0398}, and numerically intensive computations have had to be performed for $B_5$ and beyond.
(for example, $B_{10}$ requires the evaluation of 4,980,756 Ree-Hoover diagrams \cite{JSP_2006_122_0015}).
In Fig. 1 the equation of state derived using the virial coefficients up to $B_4$ is compared to the almost `exact' Carnahan-Starling equation of state \cite{JCP_1969_51_00635}.
\begin{figure}
\centering
\includegraphics[width=0.6\textwidth]{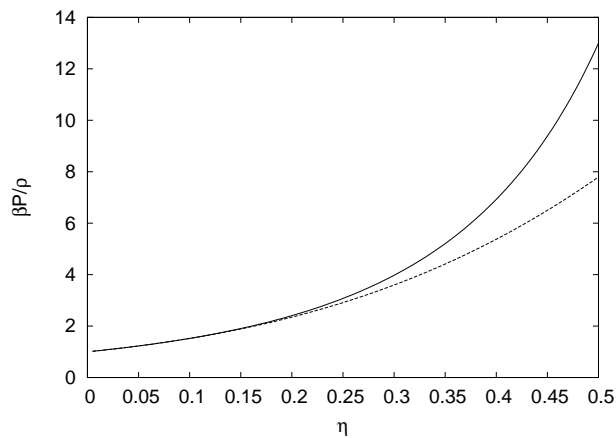}
\caption[]{
Hard sphere equation of state: the solid line Carnahan-Starling equation of state, the dashed line is the virial expansion up to $B_4$.}
\end{figure}
It can be seen that the curves start to diverge at packing fractions as low as  $\eta=0.2$
($\eta = \frac{\pi}{6} \rho d^3$, $d$ is the hard sphere diameter. The fluid-solid transition is located at $\eta \approx 0.49$).
Thus the virial equation approach is only really suitable for dilute systems and gasses.

In view of this virial expansions for liquids should be treated with caution; for three dimensional hard spheres the radius of convergence of the virial series is still 
unknown even after years of study. As yet the virial expansion is unable to reproduce the freezing transition found in simulation studies \cite{JCP_1957_27_01207,JCP_1957_27_01208}.
All said and done it should be borne in mind that the virial expansion is of empirical origin.
%%%%%%%%%%%%%%%%%%%%%%%%%
\subsection{Liquids as a disordered solid}
%%%%%%%%%%%%%%%%%%%%%%%%%
The Lennard-Jones-Devonshire cell model \cite{PRSA_1937_163_0053,PRSA_1938_165_0001,JPC_1937_41_00249} 
was based on the idea that molecules are trapped in a lattice of cages, each molecule having its own fixed cage within which it can
move freely. 
At low temperatures this cage, or well, has the parabolic form characteristic of high temperature solids. 
This well becomes increasingly an-harmonic as the  temperature is increased.
The problem with this model is that the positions of the atoms in a fluid are correlated, leading to a dynamical feedback in which
each of the atoms forming the cage has its own cage. Treating the cage as being immobile only works as a first approximation.
Cell theory is at its best near the triple point when the molecules are highly confined \cite{JCP_1950_18_01484}, its usefulness 
declining as one approaches the region of the critical point, $\rho_c$  \cite{JCP_1950_18_00380}.
%However John Kirkwood \cite{JCP_1950_18_00380} called into question the usefulness of cell theory in the absence of a solution to the 'communal entropy' problem. 

The (Lenz-)Ising model is another solid-like model \cite{PZ_1920_21_0613_nolotengo,AHES_2005_59_0267} in that it is based on a fixed regular lattice.
Good use was made of this model in  the study of critical points, 
thanks to the mathematical {\it tour de force} performed by Lars Onsager, who found an analytical solution to the 2-dimensional case \cite{PR_1944_065_000117}. 
However, it has recently been shown that an analytical solution to the 3-dimensional case is intractable \cite{P_ACM_2000_087_prEprint}.
Ising models consist of a series of lattice points. In the original implementation the lattice points were occupied by either 
spin-up or spin-down sites to represent magnets, in order to study ferromagnetism.
The so called `lattice gas' of Lee and Yang was an adaptation of the Ising model, where the spins up and down were 
replaced by occupied or un-occupied sites \cite{PR_1952_087_000404,PR_1952_087_000410}.
The  bond-fluctuation model, also based on a lattice,  has been very fruitful in the study of polymer melts \cite{MM_1988_21_2819,MTS_2003_12_0237}.
Another lattice model is that of Lebwhol and Lasher \cite{PRA_1972_06_000426} who studied the Maier-Saupe mean field model of the isotropic-nematic
transition \cite{ZN_1959_14_0882_nolotengo,ZN_1960_15_0287_nolotengo}.
However, given the analytical intractability of these models, useful solutions must be obtained via computer simulation studies.
%%%%%%%%%%%%%%%%%%%%%%%%%
\section{The advent of electronic computers}
%%%%%%%%%%%%%%%%%%%%%%%%%
From the very outset, electronic computers have been applied to the study of statistical mechanics.
Perhaps the paterfamilias of modern computers is
the `Mathematical Analyzer, Numerical Integrator, and Computer' (MANIAC-I, 1952-1956)
built by Nicholas Metropolis's  Theoretical Division at the Los Alamos National Laboratory \cite{LAS_1986_14_0096,PoP_2005_12_057303}.

MANIAC was used extensively by Marshall and Arianna Rosenbluth.
Their work included Monte Carlo simulations of two-dimensional Lennard-Jones particles and
the equation of state for three dimensional hard spheres \cite{JCP_1954_22_00881} (also examined by Alder, Frankel and Lewison \cite{JCP_1955_23_00417})
and polymer chains \cite{JCP_1955_23_00356} as well as the incredibly important Metropolis algorithm \cite{JCP_1953_21_01087}.
MANIAC-I was also used to study ergodicity by Fermi, Pasta and Ulam \cite{LA_1955_1940_nolotengo}, leading to the long-standing `FPU' problem that adopts their initials \cite{Chaos_2005_15_015106}.

In 1957, using an IBM 704 computer, Wood and Jacobson recalculated the hard sphere equation of state \cite{JCP_1957_27_01207,LASLR_1963_2827},
finding agreement with the molecular dynamics work of Alder and Wainwright \cite{JCP_1957_27_01208}.
In the same year  Wood and Parker also undertook a study of three dimensional Lennard-Jones particles \cite{JCP_1957_27_00720}.
These publications opened the door to the study of liquids from an atomistic/molecular perspective.
It is for this reason that Temperley wrote ``A key year for liquid state physics was 1957" \cite{bookTemperley}.
For personal perspectives on the early history of molecular dynamics and Monte Carlo written by some of the main
protagonists, see articles written by Wood \cite{Wood_Early_Hist} and by Rosenbluth \cite{AIPCP_2003_690_0022}.
%%%%%%%%%%%%%%%%%%%%%%%%%
\subsection{Monte Carlo and molecular dynamics}
%%%%%%%%%%%%%%%%%%%%%%%%%
The two principal techniques that are the mainstay of applied statistical mechanics of liquids are Monte Carlo integration (MC) and molecular dynamics (MD) simulations.
Stochastic methods have a long history. However, MC in its present form has its roots in the work of Nicholas Metropolis and Stanislaw Ulam \cite{JASA_1949_44_0335,ulam_1951_photocopy},
along with the Metropolis importance sampling algorithm developed by Marshall Rosenbluth (this paper having over 9300 citations at the time of writing) \cite{JCP_1953_21_01087}.
Molecular dynamics was developed a little later by Bernie Alder along with Tom Wainwright \cite{JCP_1957_27_01208}.

Classical molecular dynamics plays out the trajectories of the particles.
\begin{equation}
\frac{\partial q_i (t)}{\partial t} = \frac{ \partial q_i (\{q(t)\},\{p(t)\})}{\partial t},
\end{equation}
\begin{equation}
\frac{\partial p_i (t)}{\partial t} = \frac{ \partial p_i (\{q(t)\},\{p(t)\})}{\partial t},
\end{equation}
where $q_1,...,q_N$ are the positions of the atoms, and $p_1,...,p_N$ are the respective momenta for the $N$ degrees of freedom.
One of the most famous algorithms for solving these equations is the Verlet scheme \cite{PR_1967_159_000098}
\begin{equation}
{\bf r}_i (t + \delta t) = 2 {\bf r}_i (t) - {\bf r}_i (t -\delta t) + \delta t^2 \frac{{\bf F}_i(t)}{m_i},
\end{equation}
where $\delta t$ is the so called `time-step'.
The time step is governed by the time scale of the fastest motions in the system (for example, to about $1/10^{\rm th}$ of the bond stretching
frequency of the lightest atoms in a simulation of a molecule).

The time average of a particular observable of the system, $O$ is defined as
\begin{equation}
\overline{O}_T (\{q_0(t)\},\{p_0(t)\}) \equiv \frac{1}{T} \int_0^T O (\{q(t)\},\{p(t)\}) ~dt.
\end{equation}
These two techniques have both their advantages and disadvantages. For example, with MC it is very difficult to obtain dynamical information.
With MD phase space exploration takes much longer. 
Monte Carlo is usually performed in the canonical ensemble, leading to $p(N,V,T)$ and $U(N,V,T)$.
Molecular dynamics in the micro-canonical ensemble, giving $p(N,V,U)$ and $T(N,V,U)$.

The Ergodic hypothesis \cite{PNAS_1931_17_00656,LAS_1987_15_0263} essentially states that an ensemble average (MC) of an observable, $\left<O\right>_\mu$  is equivalent to the time average, $\overline{O}_T$ of an observable (MD).
{\it i.e.}
\begin{equation}
\lim_{T \rightarrow \infty} \overline{O}_T (\{q_0(t)\},\{p_0(t)\}) = \left<O\right>_\mu.
\end{equation}
A restatement of the ergodic hypothesis is to say that all allowed states are equally probable.
With this in mind one chooses the computational technique most suited to type of information one wishes to obtain. 
%%%%%%%%%%%%%%%%%%%%%%%%%
\subsection{Random number generators}
At the heart of any Monte Carlo computer code lies a {\it pseudo-}random number generator (RNG).
The generator is {\it pseudo} since the series of numbers originate from a deterministic computer code.
One of the first RNG's was the so called `linear congruential generator' (LCG),
proposed by D. H. Lehmer in 1951
\cite{ranLehmer51_photocopy}
%,ranMaclarenMarsaglia,CACM_1966_09_0432,CACM_1968_11_0757,ranLewisGoodmanMiller}.
The endearing features of this algorithm is that it is
simple, easy to program, portable across computing platforms and fast.
This Lehmer algorithm has come to be known as a {\it minimal standard}
against which many new algorithms are compared with.
A good number of Monte Carlo integration computer codes use the
Lehmer algorithm (or a combination thereof \cite{CACM_1968_11_0757,CACM_1988_31_0742})
to provide the pseudo-random numbers they require.
The Lehmer algorithm can be written as
\begin{equation}
y_{n+1}\equiv ay_n + b~~~({\rm mod} ~m),
\end{equation}
where the user chooses $a$, $b$, $m$, and a seed value to initiate
the algorithm, $y_0$.
A very popular implementation of the LCG is the {\it prime modulus
multiplicative linear congruential generator} (PMMLCG) \cite{CACM_1988_31_1192}. 
The parameter $m$
should be prime and as large as possible without causing a numerical overflow
on the computer that it is running on.
For example, for a 32-bit (31 bit + 1 sign bit) word size
then the logical choice of $m$ is the Mersenne prime
\begin{equation}
m=2^{31} -1=2147483647,
\end{equation}
with  $a=7^5$ (a positive primitive root of $m$ \cite{CACM_1966_09_0432,IBM_1969_02_0136}),
and $b=0$
With these parameters one is able to generate a series of
$2.147 \times 10^9$
pseudo-random numbers from one seed value.
For an interesting  discussion on how  to choose an initial seed value see \cite{CACM_2003_46_0090}.
For a list of other values of $m$ and $a$ see Ref. \cite{MC_1999_68_249}
and for its use on 64-bit computers see \cite{CPC_1997_103_0103}.

Not all random number generators
are equally good. 
\cite{CACM_1988_31_1192,MCS_1998_46_0485,PRL_1992_69_003382},
Some generators were, and actually are, quite bad
(such as the infamous RANDU \cite{CACM_1988_31_1192}).
Finding a suitable RNG for a given application is a very important task.

Given the extensive use of the Monte Carlo technique in elementary particle physics calculations
Martin L\"uscher developed a RNG called RANLUX
\cite{CPC_1994_79_0100,CPC_1994_79_0111}
(available from the CPC Program Library \cite{CPC_web})
based on the RCARRY algorithm proposed by Marsaglia and Zaman \cite{AAP_1991_01_0462}.
L\"uscher describes his algorithm as being a `discrete approximation to a chaotic dynamical system'.
The RANLUX routine has a period of an incredible  $\approx 1\times10^{171}$ random numbers.\\
%%%%%%%%%%%%%%%%%%%%%%%%%
\subsection{Liquids as a liquid}
%%%%%%%%%%%%%%%%%%%%%%%%% 
\subsubsection{Perturbation theory}
Perturbation theories are a very useful tool in many areas of physics. The idea is to take a known `simple' system as the
reference system (for example, the hard sphere fluid \cite{PR_1968_165_000201} is an ideal reference system for the study of, for example, liquid argon,
as demonstrated by Zwanzig \cite{JCP_1954_22_01420}) 
and then treat, say, attractive forces as a perturbation of the reference system. 
A classic example of a perturbation theory is the Van der Waals model \cite{bookRowlinson,RSEQ_2005_101_0019_prEprint}.
Another famous example is that of Barker and Henderson, who used the hard sphere fluid as the reference potential in a study of the 
square well system \cite{JCP_1967_47_02856}.
They also examined the 6:12 potential \cite{JCP_1967_47_04714} using the Zwanzig formulation.
Weeks, Chandler and Andersen divided the Lennard -Jones potential into repulsive and attractive sections
\cite{PRL_1970_25_000149,JCP_1971_54_05237,JCP_1971_55_05422}.
These approaches have worked well because repulsive forces are very important in fluid phase  \cite{JCP_1971_54_05237},
although naturally, without attractive forces there is no gas-liquid transition.

For `polymers' 
Wertheim developed the `first order thermodynamic perturbation theory' (TPT1) \cite{JCP_1987_87_07323,JSP_1984_35_0019_nolotengoSpringer,JSP_1984_35_0035_nolotengoSpringer,JSP_1986_42_0459_nolotengoSpringer,JSP_1986_42_0477_nolotengoSpringer}, 
 also known as the `self associating fluid theory' (SAFT) \cite{MP_1988_65_0001,MP_1988_65_1057,FPE_1989_52_0031}.
TPT1 allows one to calculate the properties of a fluid chains simply from a knowledge of the monomer fluid,
\begin{equation}
\label{tpt1}
Z_{\rm TPT1}= \frac{P}{\rho k_B T}=mZ_{\rm mon} - (m-1) \left( 1 + \rho^{\rm ref} \frac{\partial \ln g(\sigma)}{\partial \rho^{\rm ref}} \right),
\end{equation}
where $Z_{\rm mon}$ is the reference equation of state  for the monomer system, and $m$ is the number of monomers in the chain.
For example, if one wishes to study the equation of state of a fluid composed of Lennard-Jones chains one can do so
easily if one has access to the equation of state of the Lennard-Jones monomer fluid.
It is interesting to note that in this case that the TPT1 theory, originally developed for the fluid phase, has been successfully transfered to the 
solid phase \cite{JCP_2001_114_10411,JCP_2002_116_01757,JCP_2002_116_07645,MP_2003_101_2241}.

Needless to say, perturbation theories intimately depend on the quality of the data for the reference system, which 
is almost always obtained from computer simulation data.

%%%%%%%%%%%%%%%%%%%%%%%%%
\subsubsection{Integral equations}
%%%%%%%%%%%%%%%%%%%%%%%%%
In the treatment of a liquid as a liquid the Ornstein-Zernike relation \cite{KNAW_1914_17_0793} is a shining (or maybe one should say `opalescent') example.
The Ornstein-Zernike relation was born of a study of the {\it enfant terrible } of the liquid state;
a singularity known as the the critical point, found at the very end of the liquid-gas transition line.

For a homogeneous and isotropic fluid the Ornstein-Zernike relation is given by
\begin{equation} 
h({\bf r})  = c({\bf r}) + \rho \int  h({\bf r'})~c(|{\bf r} - {\bf r'}|) {\rm d}{\bf r'},
\end{equation} 
where $h(r)$ is the {\it total} correlation function (which is $g(r)$ less the mean field contribution of 1). This relation defines the new function, $c(r)$, known as the {\it direct} correlation function.
This relation expresses the notion that the `total' correlation between atoms 1 and 2 at a distance $r$ is given by 
the `direct' correlation between 1 and 2, along with indirect correlations due to contributions from
the rest of the atoms in the fluid, i.e. the influence of atom 3 on atom 2 due to the influence of atom 1 on atom 3 {\it etc.} 
This can be seen by expanding the above expression:
\begin{eqnarray}
h({\bf r}) = c({\bf r})  &+& \rho \int c(|{\bf r} - {\bf r'}|)  c({\bf r'})  {\rm d}{\bf r'} \nonumber \\
&+& \rho^2  \int \int  c(|{\bf r} - {\bf r'}|)   c(|{\bf r'} - {\bf r''}|)  c({\bf r''})   {\rm d}{\bf r''}{\rm d}{\bf r'} \nonumber \\ 
&+& \rho^3 \int\int\int  c(|{\bf r} - {\bf r'}|) c(|{\bf r'} - {\bf r''}|) c(|{\bf r''} - {\bf r'''}|) c({\bf r'''})   {\rm d}{\bf r'''}{\rm d}{\bf r''}{\rm d}{\bf r'}  \nonumber \\
%&+& \rho^4 \int \int\int\int  c(|{\bf r} - {\bf r'}|) c(|{\bf r'} - {\bf r''}|) c(|{\bf r''} - {\bf r'''}|) c(|{\bf r'''} - {\bf r''''}|) h({\bf r''''})  {\rm d}{\bf r''''} {\rm d}{\bf r'''}{\rm d}{\bf r''}{\rm d}{\bf r'} \\
&+& ...
\end{eqnarray}
Obviously, a fluid, because of its nature, is composed of a large number of atoms. This makes the aforementioned exact relation 
unwieldy and impracticable to say the least.
For this under-determined equation to be useful one must look for a closure relation. Such closure relations are either asymptotic schemes, 
involving much mathematical work but leading to an estimate of the errors involved, or some form of truncation scheme, in which higher order moments are arbitrarily
assumed to vanish, but  possibly resulting in a drastic approximation.
In 1960 Morita and Hiroike \cite{PTP_1960_023_1003}
produced the formally exact closure formula,
\begin{equation}
h(r) -1= \exp\left(-\beta \Phi(r) + h(r) -c(r) +B[h(r)]\right), 
\end{equation}
where $\Phi(r)$ is the pair potential.
This reduced the task to finding the so called `bridge' functional $B[h(r)]$ (so named due to its diagrammatic similarity to the famous Wheatstone bridge)
rather than the entire closure relation.
However, the bridge functional represents an infinite sum of diagrams. 
Sadly, the bridge function is not known exactly even for the simplest model;
one component hard spheres. Apart form the most simple cases (for example, for the Percus-Yevick closure \cite{PR_1958_110_000001} for hard spheres \cite{PRL_1963_10_000321,JMP_1964_05_00643} or mixtures thereof \cite{PR_1964_133_00A895}) 
modern use of integral equations have to rely on computers to 
obtain iterative solutions \cite{MP_1979_38_1781,MP_1985_56_0709,MP_1989_68_0087} to the various approximate closures of the OZ equation.
%%%%%%%%%%%%%%%%%%%%%%%%%
\section{Progress}
%%%%%%%%%%%%%%%%%%%%%%%%%
%%%%%%%%%%%%%%%%%%%%%%%%%
%%%%%%%%%%%%%%%%%%%%%%%%%
In this section I shall focus on a small number of liquid systems that are currently benefitting, and will continue to benefit from, 
fast electronic calculating machines.
The original simulation of 56 two-dimensional Lennard-Jones atoms was performed in 1953 \cite{JCP_1954_22_00881}
on the MANIAC machine at Los Alamos. In 2003 a simulation was performed, on the QSC machine again at Los Alamos, 
for a system of 19,000,416,964 splined Lennard-Jones sites \cite{IJMPC_2004_15_0193}.
In terms of a Moore's Law \cite{E_1965_38_Moore} like scaling, 
from these two data points the number of atoms that one can simulate seems to double roughly every 21 months.
Thus any computational {\it tour de force} mentioned here is perhaps by now pass\'e.
%%%%%%%%%%%%%%%%%%%%%%%%%
\subsection{First Principles Molecular Dynamics}
%%%%%%%%%%%%%%%%%%%%%%%%%
First principles molecular dynamics, developed by Car and Parrinello \cite{PRL_1985_55_002471,SSC_1997_102_0107},
involves solving the Kohn-Sham equations \cite{PR_1965_140_0A1133}. These are a set of non-linear, coupled integro-differential partial differential equations
which make use of the Born-Oppenheimer approximation \cite{AdPL_1927_84_0457_nolotengoWiley} (for a recent study 
of the validity of this approximation see \cite{S_2002_296_00715}).
Thus ions can be treated as classical particles obeying Newton's laws of motion.
The potential is defined as 
\begin{equation} 
\Phi(r_i) = \left< \Phi_0 \vert \mathcal{H}(r_i) \vert \Psi_0 \right>,
\end{equation}
where $ \mathcal{H}(r_i)$ is the many-body electronic Hamiltonian, and $\Psi_0$ is the ground state eigenfunction.
The forces are then found using the Hellman-Feynman theorem \cite{PR_1939_056_000340},
\begin{equation} 
\frac{\partial \Phi(r_i)}{\partial r_i} = \left< \Psi_0 \left \vert \frac{\partial \mathcal{H}(r_i)}{\partial  r_i} \right \vert \Psi_0 \right>.
\end{equation}
One of the beauties of first-principles molecular dynamics (FPMD) is that it is able to simulate chemical reactions, {\it i.e.} the making and breaking of
chemical bonds, something not open to standard molecular dynamics techniques.
On the other hand, one of the great hindrances to FPMD is that the simulations 
scale as $N^3$ where $N$ is the number of ions.
Here I should like to mention a recent simulation, not of a liquid but of a solid metal. I do this because of the impressive size of the system:
1000 atoms of the transition metal molybdenum  (Mo) \cite{Qbox_web} (a system comprising of 12,000 electrons) have been studied 
on the Blue Gene/L computer, currently the worlds fastest computer  (having 131,072 processors).
Another system, studied using pseudopotential theory interaction potentials, is the solidification of 32,768,000 atoms of 
the molten metal tantalum (Ta) \cite{PRL_2006_96_225701}, again using the Blue Gene/L computer.
%%%%%%%%%%%%%%%%%%%%%%%%%
\subsection{Liquid Crystals}
%%%%%%%%%%%%%%%%%%%%%%%%%
Liquid crystals are a phase of matter situated between the liquid and the solid phases. They flow as liquids do,
but also present crystalline-like properties, such as birefringence ({\it i.e.} are anisotropic).
Liquid crystals were discovered in 1888 by the biochemist Friedrich Reinitzer \cite{MfC_1888_9_421_nolotengoSpringer,CR_1957_57_01049,bookCrystalsFlow}
whilst studying the compound cholesteryl benzoate obtained from carrots. 
This compound was further and extensively studied by Otto Lehmann \cite{ZfPC_1889_4_0462}.
In a classic treatise, in 1922, Georges Friedel recognised liquid crystals as comprising a new form of matter, 
and reluctantly coined new names for the then known liquid-crystalline, or {\it mesomorphic} phases;
nematic ($\nu \acute{\eta} \mu \alpha$, thread), smectic ($\sigma \mu \acute{\eta} \gamma \mu \alpha$, soap), and cholesteric \cite{AP_1922_18_273,DFS_1958_25_0019}.
Nematic liquid crystals have orientational order, but lack positional order (see Fig. 2). Smectic phases form a layered structure, and although 
they too also have orientational order, they lack positional order {\it within} the layers.
Today, liquid crystals are ubiquitous as the active component of a whole spectrum of display devices.

One of the distinguishing feature of liquid crystals is their shape anisotropy.
Lars Onsager \cite{ANYAS_1949_51_0627_nolotengo} showed that (lyotropic) nematic phases form at vanishingly low 
concentrations for a solution of infinitely thin, infinitely long, rods.
He derived an expression for the free energy in terms of an orientational distribution function,
where the  orientational and positional contributions can be considered independently.
In sacrificing orientational entropy ({\it  i.e.} by aligning) many more positions are freed up.
This competition eventually leads to an orientationally ordered phase as the concentration, or density, is increased.
However, an understanding of the liquid crystal phase for 
less idealised models had to wait for the advent of computer simulations.
One of the first simulations was that of Vieillard-Baron \cite{JCP_1972_56_04729}, who, in 1972, 
studied the I-N transition for a system of 170 two-dimensional hard ellipses.
\begin{figure}
\centering
\includegraphics[width=0.6\textwidth]{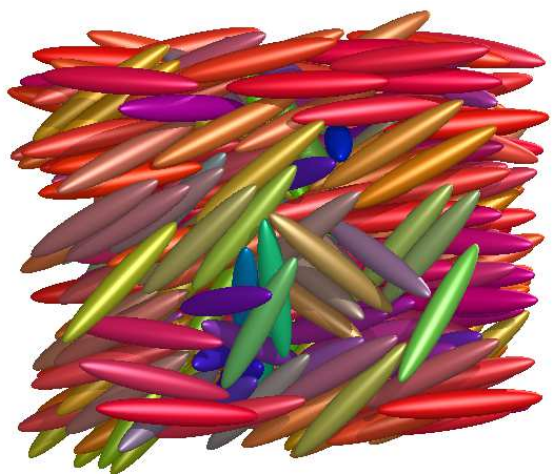}
\caption[]{
Snapshot of a   nematic liquid crystal phase from a Monte Carlo simulation of hard prolate  ($1\times1\times6$) ellipsoids.
}
\end{figure}
A wonderful simulation was the growth of a smectic phase in 1988 by Frenkel, Lekkerkerker and Stroobants \cite{N_1988_332_00822}
for a system of 576 hard spherocylinders having a length to diameter ratio of 5.
For the case of linear-tangential hard spheres, Monte Carlo simulations have produced  smectic phases for
$m=5$ and both nematic and smectic phases for $m\geq 6$  \cite{JCP_2001_115_04203} .
The first demonstration of spontaneous liquid crystal phase formation for a `realistic' mesogen 
was shown by McBride, Wilson and Howard in 1998 \cite{MP_1998_93_0955} in a molecular dynamics simulation
of the molecule 4,4'-di-n-pentyl-bibicyclo [2.2.2]octane.
In 2001 the first thousand molecule simulation of a mesogen at the fully atomistic level
was performed \cite{MCLC_2001_363_181_prEprintTF} for the molecule PCH5.

For two excellent reviews on liquid crystal simulations see Wilson \cite{IRPC_2005_24_0421} and  Care and Cleaver \cite{RPP_2005_68_2665}.
%%%%%%%%%%%%%%%%%%%%%%%%%
\subsection{Water}
%%%%%%%%%%%%%%%%%%%%%%%%%
Water has elicited sustained scientific interest for well over 100 years, notably the early work of Gustav Tammann and Percy Bridgman 
\cite{AdPL_1900_2_0001_nolotengoWiley,PAAAS_1912_XLVII_13_441_photocopy,PNAS_1915_1_00513,JCP_1935_03_00597,JCP_1937_05_00964}. 
This is partly because of its `anomalous' physical properties, and partly because of its integral 
role in biological processes.
Water still holds many surprises install, with ices XIII and XIV having only just been found in 2006 \cite{S_2006_311_01758,JCP_2006_125_116101}.
Pioneering simulation work was performed by
Barker Watts (Monte Carlo) \cite{CPL_1969_03_0144}
and Rahman and Stillinger (molecular dynamics) \cite{JCP_1971_55_03336}.
Since then a great many `simple' models of water have been proposed. For an extensive review see Guillot  \cite{JML_2002_101_0219}.
The most popular models seem to be the 
`four point transferable intermolecular potential' (TIP4P) \cite{JCP_1983_79_00926} (with over 5500 citations at the time of writing),
and the `simple point charge-extended' (SPC/E) model \cite{JPC_1987_91_06269} (with nearly 2000 citations at the time of writing).
The more recent TIP5P \cite{JCP_2000_112_08910} is also attracting substantial interest. 
A wonderful example of the power of such seemingly simple models is shown when one calculates the 
water and ice phase diagram.
This task was recently undertaken by the Vega group for the TIP4P and SPC/E models \cite{PRL_2004_92_255701}
using a combination of Monte Carlo simulations along with Gibbs-Duhem integration \cite{JCP_1993_98_04149}
to trace out the phase boundaries.
The TIP4P model is able to make a fair representation of the experimental phase diagram. Not at all bad
if one considers that the model was designed solely for the liquid phase.
Water has been studied up to 70,000 K and 3.7 g/cm$^3$ using density functional theory \cite{PRL_2006_97_017801},
and it's structure has been examined using FPMD (see \S 4.1) in the complete basis set limit \cite{JCP_2006_125_154507} 
resulting in good agreement with experimental results obtained from neutron scattering.
%%%%%%%%%%%%%%%%%%%%%%%%%
\subsection{Biomolecular systems}
%%%%%%%%%%%%%%%%%%%%%%%%%
Biomolecular systems are composed of one or various macromolecules, often immersed in 
a solvent, such as water. 
However, their size is not the only obstacle to simulations of such systems.
Proteins, in their {\it denatured} state adopt a more-or-less random configuration.
Proteins in their {\it natural} state almost un-erringly 
adopt  a unique conformation (tertiary structure). However, this is usually 
just one local minima of potentially a large number of minima.
The essential question is how do these proteins repeatedly and reversibly `fold' into their
biologically active conformation. 
Given the size of proteins, an exploration of phase space in order to arrive at 
a minimum energy conformation would take an inordinate amount of time; however, 
protein folding is rapid ($\approx 100$ns for $\alpha$ helices, 1$\mu$s for $\beta$ hairpins \cite{ACR_1998_31_0745}) in an adequate (physiological) environment.
This led to the idea of a {\it pathway} for protein folding.
These questions were raised in two  summaries  written by Cyrus Levinthal \cite{JCP_PCB_1968_65_0044_photocopyEDP,Levinthal_nolotengo}
and resulted in a Nobel Prize in chemistry for Christian Anfinsen \cite{S_1973_181_00223} for his
`thermodynamic hypothesis'; the Gibbs free energy of the system is at a minimum in the natural state.

A land-mark study was the 1 microsecond simulation of the 36-residue villin headpiece subdomain 
by Duan and Kollman \cite{S_1998_282_00740,ARPC_2000_51_0435}. Although 1 microsecond 
was not sufficient to see the entire folding process (of the order of 10-100 $\mu$s) 
it demonstrated the feasibility of such simulations.
Recently, an even smaller 
protein, an artificial peptide called chignolin, has been synthesised. It consists of only
10 amino acid residues \cite{Str_2004_12_1507}. This peptide  has been the subject of a replica exchange molecular dynamics (REMD) simulation \cite{JMB_2005_354_0173}
leading to reproducible folding.
Hubner, Deeds, and Shakhnovich have since developed  a fully transferable potential  for use in all-atom models
\cite{PNAS_2005_102_18914}
that results in reliable `high-resolution' folding.
It is also worth noting that although these biomolecules are very large, it is not unusual to 
dedicate a substantial part of the computation task to the simulation of the solvent  \cite{CR_2000_100_04187}.
%Complete satellite Tobacco Mosaic Virus  1 million atoms \cite{Str_2006_14_0437}
%Coarse graining \cite{COSB_2005_15_0144,BJ_2005_89_2372}
%Review solvent effect \cite{CR_2000_100_04187}
%Molecular crowding \cite{PNAS_2005_102_04753}

All said and done, there are occasions where the incredibly complex folding process goes awry and the proteins misfold \cite{N_2003_426_00884}, 
either following their synthesis on the ribosome,
or later, resulting in harmful isoforms known as prions  \cite{PNAS_1993_90_10962,COSB_1997_7_0053}.
Misfolded proteins tend to form aggregates whose deposits can lead to devastating conditions such as 
Alzheimer's and Parkinsons diseases \cite{N_2003_426_00900}.
Thus a good understanding of how proteins fold is one of the most important present-day challenges.
A testament to this importance is that IBM has built the worlds currently fastest supercomputer
(the aforementioned Blue Gene/L machine) with these problems in mind.

It is important to mention the Folding@Home distributed computing initiative \cite{Folding_web}. 
This project makes use of the millions of computers in the world that are lying idle at any one point in time.
Members of the public are free to download mini-simulations, run them as a screen-saver, and then return 
the completed calculations to Folding@Home.

%%%%%%%%%%%%%%%%%%%%%%%%%
\section{The Future}
%%%%%%%%%%%%%%%%%%%%%%%%%
To prognosticate about the future is a notorious business.
However, in the short term wonderful new insights into the liquid state are inevitable.

The invention of the World Wide Web at the Conseil Europ\'{e}en pour la Recherche Nucl\'{e}aire (CERN) 
in 1989 revolutionised the distribution and communication
of information. 
With over 90\% of all the scientific papers ever published now available on-line, one can perform
a comprehensive literature search in maybe less than an hour. 

Now, once again, CERN is the driving force behind  another change in the way we work.
The development of the biggest physics experiment ever, the Large Hadron Collider (LHC),
will result in an incredible amount of data in need of analysis.
It is estimated that when in operation the LHC will produce an overwhelming 15 million gigabytes of information per year.
In order to process such a vast amount of data CERN is supporting the 
LHC Computing Grid and the Enabling Grids for E-sciencE projects. 
A GRID is, as its name suggests, a grid of computers connected via very high speed Internet connections
(around one gigabyte/second).
The idea is to share computing power and storage capacity on a global scale, thus effectively
creating one vast computational resource.\\

Data, usually encountered in a plethora of proprietary formats, can be difficult
to analyse or visualise without specialised programs. It is likely that some format based on the 
W3C Extensible Markup Language (XML \cite{XML_web}) will allow greater access to this data,
and will result in more efficient data management and analysis.

On a local scale, open software, such as the Linux operating system, 
along with off-the-shelf hardware have allowed research groups
to build mini-super computers at relatively little cost.
A model for such systems are the so-called Beowulf clusters \cite{beowulf_web},
using the Message Passing Interface (MPI) \cite{MPI_web} calls for parallel computer programs.

Simulations that can be performed independently, or `concurrently' 
have found popular support. The enthusiasm of the general public for 
science is amply demonstrated by distributed computing `@' projects, such as the hugely 
successful Folding@Home, having access to nearly two million home computers with 
an estimated 200 TFLOPS of collective computing power at the time of writing. 

50 years on from the first computer simulations on MANIAC-I, Los Alamos National Laboratory will be host to the
IBM `Roadrunner' super-computer, with a projected sustained speed of 1 petaflop and a peak performance of 1.6 petaflops.
Roadrunner will have over 32,000 Cell/AMD Opteron processors, a long way from the 1024 vacuum tubes of the MANIAC-I.\\
Could 2007 be another key year for liquid state physics?
%%%%%%%%%%%%%%%%%%%%%%%%%
\section{Acknowledgements}
%%%%%%%%%%%%%%%%%%%%%%%%%
I would like to thank the Spanish CSIC for the award of an I3P grant,
as well as  project FIS2004-02954-C03-02 of the Spanish
Ministerio de Educacion y Ciencia, and project S-0505/ESP/0299 - CSICQFT
(MOSSNOHO) of the D. G. de Universidades e Investigaci\'on del Comunidad de Madrid.
I also would like to thank Enrique Lomba for a critical reading of this manuscript.
\bibliography{bibliography,local}
\end{document}